\documentclass[prc,twocolumn,superscriptaddress,superscriptfootnote]{revtex4}
\usepackage{amssymb}

\usepackage{amsmath,amsfonts,amssymb,epsfig,color}

\begin{document}
\title{Phase Structure of the Interacting Vector Boson Model}
\author{H. G. Ganev}
\affiliation{Institute of Nuclear Research and Nuclear Energy,
Bulgarian Academy of Sciences, \\ Sofia 1784, Bulgaria}

\setcounter{MaxMatrixCols}{10}

\begin{abstract}
The two-fluid Interacting Vector Boson Model (IVBM) with the $U(6)$
as a dynamical group possesses a rich algebraic structure of
physical interesting subgroups that define its distinct exactly
solvable dynamical limits. The classical images corresponding to
different dynamical symmetries are obtained by means of the coherent
state method. The phase structure of the IVBM is investigated and
the following basic phase shapes, connected to a specific geometric
configurations of the ground state, are determined: spherical,
$U_{p}(3)\otimes U_{n}(3)$, $\gamma-$unstable, $O(6)$, and axially
deformed shape, $SU(3)\otimes U_{T}(2)$. The ground state quantum
phase transitions between different phase shapes, corresponding to
the different dynamical symmetries and mixed symmetry case, are
investigated.
\end{abstract}
\maketitle PACS number(s): {21.60.Ev, 21.60.Fw, 21.10.Re}

\section{Introduction}

The phase structure of quantum many-body systems has been a subject
of great experimental and theoretical interest in the last years.
The introduction of the concept of critical point symmetry
\cite{CPS} has recalled the attention of the community to the topic
of quantum phase transitions in nuclei. Different models have been
used to describe the quantum phase transitions in different
many-body systems, such as atomic nuclei \cite{IBM}, molecules
\cite{VM},\cite{molecules}, atomic clusters \cite{atomclust}, and
finite polymers. Among these models, those based upon algebraic
Hamiltonians play an important role.

There are many approaches which allow the association of a certain
geometry to any abstract algebra, but for algebraic models, this can
be achieved with the theory of the coherent states
\cite{MF1},\cite{MF2},\cite{MF3},\cite{ZFG}. The expectation value
of the Hamiltonian in the ground coherent state is refereed to as
its classical limit. The classical limit of quantum systems is one
of the oldest problems in quantum mechanics. This problem appears
whenever one formulates a theory in terms of quantum variables and
wishes to interpret it in terms of classical (geometrical)
variables. The method of coherent (or intrinsic) states provides a
prescription for translating algebraic operators into canonical
phase-space coordinates, thereby allowing algebraic models of
nuclear structure and dynamics to be interpreted from the
perspective of their corresponding classical limits. Of particular
relevance for the present study is the construction of the potential
energy or potential functions \cite{IBM},\cite{VM}.

A nice feature of the algebraic models is the occurrence of phases
connected to a specific geometric configurations of the ground
state, which arise from the occurrence of different dynamical
symmetries. The study of the ground state energy as a function of an
appropriately chosen parameter, called control parameter, shows a
transition between the different phases. These phase transitions are
referred to as ground-state or quantum phase transitions and have
been widely investigated in the last years (e.g., a review article
\cite{IBMQPT}). Since these transitions are between different
shapes, they are sometimes termed as "shape transitions".

An important aspect of the study of phase transitions is the
construction of the phase diagram (structure). In this respect it is
interesting to see what is the phase structure of the two-fluid
Interacting Vector Boson Model (IVBM) \cite{IVBM}. Thus, it is the
purpose of the present paper to investigate what are the different
phase shapes which might occur within the framework of the IVBM. The
first step in the construction of the phase diagram is the
identification of all possible dynamical symmetries of the system.
In IVBM there are several dynamical symmetries which will be
considered in next sections. It should be shown that there exist
three distinct shapes corresponding to the three dynamical
symmetries of IVBM:\quad (1) spherical shape, $U_{p}(3)\otimes
U_{n}(3)$, \quad (2) $\gamma-$unstable shape, $O(6)$, and \quad (3)
axially deformed shape, $SU(3)\otimes U_{T}(2)$.

\section{The algebraic structure generated by the two vector bosons}

The algebraic structure of IVBM \cite{IVBM},\cite{pepan} is realized
in terms of creation and annihilation operators of two kinds of
vector bosons $u_{m}^{\dag}(\alpha )$, $u_{m}(\alpha )$ ($m=0,\pm
1$), which differ in an additional quantum number $\alpha=\pm1/2$
(or $\alpha=p$ and $n$)$-$the projection of the $T-$spin (an
analogue to the $F-$spin of IBM-2). We consider these two bosons
just as building blocks or "quarks" of elementary excitations
(phonons) rather than real fermion pairs, which generate a given
type of symmetry. In this regard, the $s$ and $d$ bosons of the
IBM-1 can be considered as bound states of elementary excitations
generated by two vector bosons. Thus, we assume that it is the type
of symmetry generated by the bosons which is of importance for the
description of the collective motions in nuclei rather than the
tensorial or fermionic nature of these bosons.

The number preserving bilinear products of the creation and
annihilation operators of the two vector bosons generate the boson
representations of the unitary group $U(6)$
\cite{IVBM},\cite{pepan}:
\begin{equation}
A_{M}^{L}(\alpha, \beta )={\sum
}_{k,m}C_{1k1m}^{LM}u_{k}^{\dag}(\alpha )u_{m}(\beta ),
\label{numgen}
\end{equation}
where $C_{1k1m}^{LM}$, which are the usual Clebsch-Gordan
coefficients for $L=0,1,2$ and $M=-L,-L+1,...L$, define the
transformation properties of (\ref{numgen}) under rotations. We will
also use the notations $u_{m}^{\dag}(\alpha=1/2 )=p^{\dag}_{m}$ and
$u_{m}^{\dag}(\alpha=-1/2 )=n^{\dag}_{m}$.

In the most general case the two-body model Hamiltonian should be
expressed in terms of the generators of the group $U(6)$. In some
special cases the Hamiltonian can be written in terms of the
generators of different subgroups of $U(6)$. The $U(6)$ group
contains the following chains of subgroups \cite{IVBM},\cite{pepan}:
\begin{widetext}
\begin{equation}
\begin{tabular}{lll}
\begin{tabular}{l}
$\ \ \ \ \ \ \ \ \ $ \\
$\ \ \ \ \ \ \ \ \ \ \ \ \ \ \ \ \ \ \ \swarrow \ \ \ $ \\
$\ \ U_{p}(3)\otimes U_{n}(3)$ \\
$\ \ \ \ \ \ \ \ \ \ \ \downarrow $ \\
$SO_{p}(3)\otimes SO_{n}(3)$%
\end{tabular}
&
\begin{tabular}{l}
$\ \ \ \ \ \ \ \ \ \ \ \ \ U(6)$ \\
$\ \ \ \ \ \ \ \ \ \ \ \ \ \ \ \downarrow \ \ \ \ \ \ \ \ $ \\
$\ \ \ \ \ \ \ \ \ \ \ \ O_{\pm }(6)$ \\
$\swarrow \ \ \ \ \ \ \ \ \ \ \ \ \downarrow \ \ \ \ \ $ \\
$\ \ \ \ \overline{SU_{\pm }(3)}\otimes SO(2)$%
\end{tabular}
& \ \ \ \
\begin{tabular}{l}
$\ $ \\
$\searrow $ \\
$\ \ U(3)\otimes U_{T}(2)$ \\
$\ \ \ \ \ \ \ \ \ \downarrow $ \\
$SU_{\pm }(3)\otimes SO_{T}(2)$%
\end{tabular}
\\
\ \ \ \ \ \ \ \ \ \ \ \ \ \ \ \ \ \ \ \ \
\begin{tabular}{l}
$\searrow $ \\
\end{tabular}
& \ \ \ \ \ \ \ \ \ \
\begin{tabular}{l}
$\ \ \ \ \ \downarrow $ \\
$\ \ SO(3)$%
\end{tabular}
& \ \ \ \
\begin{tabular}{l}
$\swarrow $ \\
\end{tabular}%
\end{tabular}
\label{chains}
\end{equation}
\end{widetext}

As can be seen, the IVBM has a rich enough algebraic structure of
subgroups. Each of these dynamical symmetries will correspond to a
certain possibly different shape phase. Now we are going to briefly
enumerate these algebras, their generators and some of their Casimir
operators relevant to the present work.

\subsection{The $U_{p}(3)\otimes U_{n}(3)$ chain}

\quad (a) $U_{p}(3)\otimes U_{n}(3)$ algebra. It consists of two
sets of commuting operators:
\begin{align}
&N_{p}=\sqrt{3}(p^{\dagger}\times p)^{(0)}, \label{Np} \\
&L^{p}_{M}=\sqrt{2}(p^{\dagger}\times p)^{(1)}_{M}, \label{Lp} \\
&Q^{p}_{M}=\sqrt{2}(p^{\dagger}\times p)^{(2)}_{M}. \label{Qp}
\end{align}
and
\begin{align}
&N_{n}=\sqrt{3}(n^{\dagger}\times n)^{(0)}, \label{Nn} \\
&L^{n}_{M}=\sqrt{2}(n^{\dagger}\times n)^{(1)}_{M}, \label{Ln} \\
&Q^{n}_{M}=\sqrt{2}(n^{\dagger}\times n)^{(2)}_{M}. \label{Qn}
\end{align}
The $SU_{\tau}(3)$ ($\tau=p,n$) algebra is obtained by excluding the
number operator $N_{\tau}$, whereas the angular momentum algebra
$SO_{\tau}(3)$ is generated by the generators $L^{\tau}_{M}$ only.

For the decomposition of fully symmetric irreducible representation
(IR) $ [N]_{6}$ of $U(6)$ into the IR's of $U_{p}(3) \otimes
U_{n}(3)$, we have \cite{VanBook}:
\begin{equation}
\lbrack N]_{6}=\sum_{m=0}^{N}[N-m]_{3}\otimes \lbrack m]_{3},
\end{equation}
or in Elliott notations
\begin{equation}
\lbrack N]_{6}=\sum_{m=0}^{N}(\lambda _{p}=N-m,\mu _{p}=0)\otimes
(\lambda _{n}=m,\mu _{n}=0). \label{U6U3U3}
\end{equation}
For the decomposition $SU_{\tau}(3)\supset SO_{\tau}(3)$, the
standard reduction rules take place \cite{VanBook}
\begin{align}
&K=\min (\lambda ,\mu ),\min (\lambda ,\mu )-2,...,0~(1) \notag \\
&L=\max(\lambda ,\mu ),\max (\lambda ,\mu )-2,...,0~(1);K=0 \notag \\
&L=K,K+1,...,K+\max (\lambda ,\mu );K\neq 0. \label{su3so3}
\end{align}

The linear Casimir operators are simply
\begin{equation}
C_{1}[U_{p}(3)] =N_{p}  \label{C1p}
\end{equation}
and
\begin{equation}
C_{1}[U_{n}(3)] =N_{n}. \label{C1n}
\end{equation}
The quadratic Casimir operator of the one-fluid $U_{\tau}(3)$
algebra can be expressed in the following multipole form
\begin{equation}
C_{2}[U_{\tau}(3)] =N_{\tau}(N_{\tau}+3), \label{C2U3t}
\end{equation}
and that of $SU_{\tau}(3)$ as
\begin{equation}
C_{2}[SU_{\tau}(3)]=\frac{4}{3}Q^{\tau} \cdot
Q^{\tau}+\frac{1}{2}L^{\tau} \cdot L^{\tau}. \label{C2SU3t}
\end{equation}

\quad (b) $O_{p}(3)\otimes O_{n}(3)$ algebra. This algebra is
determined by the operators (\ref{Lp}) and (\ref{Ln}).

\subsection{The $SU(3)\otimes U_{T}(2)$ chain}

The $SU(3)\otimes U_{T}(2)$ algebra also consists of two commuting
sets of operators:

\quad (a) $U_{T}(2)$ algebra. It is defined by the operator of a
number of particles
\begin{equation}
N=N_{p}+N_{n} \label{N}
\end{equation}
and the "T-spin" operators $T^1_m,(m=0,\pm1)$ introduced through
\begin{align}
T^{1}_{1} &= \sqrt{\tfrac{3}{2}}A^0(p,n), \qquad T^{1}_{-1} = -\sqrt{\tfrac{3}{2}}A^0(n,p) \label{Trl} \\
T^{1}_{0} &= -\tfrac{\sqrt{3}}{2}\left[A^0(p,p) - A^0(n,n)\right] .
\label{T0}
\end{align}
The above operators $T^1 _m (m=0,\pm1)$ commute with $N$. Thus
(\ref{Trl}) and (\ref{T0}) define the subalgebra $su_{T}(2)\subset
u_{T}(2)$. These operators play an important role in the
consideration of the nuclear system as composed by two interacting
(proton and neutron) subsystems. At fixed $N$ the "T-spin" $T$ takes
the values $T=N/2,N/2-1,\ldots, 0$ or $1$, respectively. The IR's of
the subgroup $SO_{T}(2) \subset SU_{T}(2)$ generated by $T^{1}_{0}$
are determined standardly as follows: $-T, -T + 1, \ldots, T$. The
second-order Casimir operator of $U_{T}(2)$ is
\begin{displaymath}
C_{2}[U_{T}(2)] =\tfrac{4}{3}T^{2}+\tfrac{1}{3}N^{2}
\end{displaymath}
with eigenvalue $\frac{4}{3}T(T+1) + \frac{1}{3}N^{2}$.

\quad (b) $U(3)$ algebra. It consists of the operators which are a
sum of the $p-$ and $n-$boson subsystem operators: the total number
of bosons $N$ (\ref{N}), the angular momentum operator
$L_M=L^{p}_{M}+L^{n}_{M}$ and the components of the "truncated" (or
Elliott's) quadrupole operator
\begin{equation}
Q_M = Q^{p}_{M}+Q^{n}_{M}. \label{Qplus}
\end{equation}
The operators $L_M$ and $Q_M$ commute with $N$ and define the
subalgebra $su(3)\subset u(3)$.

The second-order Casimir operator of $U(3)$ is
\begin{equation}
C_{2}[U(3)]=\tfrac{1}{6}Q^2 +\tfrac{1}{2}L^{2}
+\tfrac{1}{3}N^{2}=\tfrac{1}{2}N^{2}+2T^{2}+N, \label{C2U3}
\end{equation}
where
\begin{displaymath}
Q^2 = 6\sum_{M,\alpha,\beta}(-1)^M A^2_M(\alpha,\alpha)
A^2_M(\beta,\beta)\ .
\end{displaymath}
The $SU(3)$ Casimir is simply
\begin{equation}
K_{3}=\tfrac{1}{6}Q^2 +\tfrac{1}{2}L^{2} \label{C2SU3}
\end{equation}
and its eigenvalue is $(\lambda^{2}+\mu^{2}+\lambda \mu
+3\lambda+3\mu)$. It should be pointed that the groups $U_{T}(2)$
and $U(3)$ are mutually complementary in the sense that the
eigenvalues of $C_{2}[U(2)]$ are completely determined by the
eigenvalues of $C_{2}[U(3)]$. This is due to the relation
\begin{equation}
C_{2}[U(3)]= \tfrac{3}{2}C_{2}[U_{T}(2)] + N . \label{C2U3}
\end{equation}
This means that the representations of $U(3)$ and $U_{T}(2)$ can be
labeled by the same quantum numbers (for instance, the number of
quasiparticles $N$ and the "T-spin" $T$).

With the help of standard group-theoretical methods \cite{VanBook}
one obtains the quantum numbers $[E_{1},E_{2},E_{3}]_{3} \times
[\varepsilon_{1},\varepsilon_{2}]_{2}$ labelling the IR of the
direct product $U(3) \otimes U_{T}(2)$ belonging to a given IR of
$U(6)$. In the case of a fully symmetric IR $ [N]_{6}$ of $U(6)$ the
decomposition is
\begin{equation}
\lbrack N]_{6}=\sum_{i=0}^{<\frac{N}{2}>}[N-i,i,0]_{3} \times
[N-i,i]_{2},  \label{U6U3U2}
\end{equation}
where $<\frac{N}{2}>=\frac{N}{2}$ if $N$ is even and $\frac{N-1}{2}$
if $N$ is odd. In fact this means that $[N]_{6}$ decomposes into
$SU(3)$ multiplets with $(\lambda,\mu)=(N-2i,i)$, $i= 1,2,\ldots,
<\frac{N}{2}>$. Each of these multiplets is characterized by a
'T-spin' $T= \frac{\lambda}{2}$. The decomposition of any IR
$(\lambda,\mu)$ of $SU(3)$ into the $SO(3)$ representations $L$ is
given by  (\ref{su3so3}).

\subsection{The $O_{\pm}(6)$ chain}

\quad (a) $O(6)$ algebra. The $O_{+}(6)$ algebra is spanned by the
following operators:
\begin{align}
&A^1_M(p,p)\equiv L^{p}_{M}, \\
&A^1_M(n,n)\equiv L^{n}_{M}, \\
&G^{(+)}_{ij}=p^{\dagger}_{i}n_{j}+n^{\dagger}_{i}p_{j}. \label{O6a}
\end{align}
An alternative $O_{-}(6)$ algebra can be defined with the generators
\begin{equation}
G^{(-)}_{ij}=\mathrm{i}(p^{\dagger}_{i}n_{j}-n^{\dagger}_{i}p_{j})
\label{O6b}
\end{equation}
instead of $G^{(+)}_{ij}$. Both the $O_{+}(6)$ and $O_{-}(6)$
algebras have the same eigenspectrum but differ through phases in
the wave functions. They are related by a transformation that is a
special case of a wider class of transformations known as inner
automorphisms. It is known that representation theory does provide
all of the embeddings, but it does not provide all of the dynamical
symmetries \cite{HS}. Indeed, the inner automorphisms can provide
new dynamical symmetry limits, sometimes referred as to "hidden"
\cite{HS} or "parameter" symmetries \cite{parsym}.

The second order Casimir operator of $O(6)$ is given by
\begin{equation}
C_{2}[O(6)]=3(G^{\pm} \times G^{\pm})^{(0)}+L^{2}_{p}+L^{2}_{n}.
\label{C2O6}
\end{equation}
and its eigenvalue $\omega(\omega+4)$ is determined by the quantum
number $\omega$ characterizing the $O(6)$ IRs.

\quad (b) $\overline{SU_{\pm}(3)}\otimes SO(2)$ algebra. The
$\overline{SU_{\pm}(3)}\otimes SO(2)$ algebra consists of two
commuting sets of operators:

\quad $\circ$  $\overline{SU(3)}$ algebra. There are two distinct
$\overline{SU_{\pm}(3)}$ algebras which are generated by
\begin{equation}
B^{(\pm)}_{ij}= G^{(\pm)}_{ij}-\frac{1}{3}\delta_{ij}M,
\label{SU3B}
\end{equation}
where
\begin{equation}
M = \sum_{i}G^{(\pm)}_{ii}. \label{SO2}
\end{equation}
Its second-order Casimir operator is
\begin{equation}
\overline{G_{3}} = \sum_{ij}B^{(\pm)}_{ij}B^{(\pm)}_{ji} .
\label{C2SU3B}
\end{equation}

\quad $\circ$ $SO(2)$ algebra. The generator of this algebra is
given by (\ref{SO2}).

For $O(6)\subset U(6)$, the symmetric representation $[N]_{6}$ of
$U(6)$ decomposes into fully symmetric $(\omega ,0,0)_{6}\equiv
(\omega )_{6}$ IR's of $O(6)$ according to the rule
\cite{VanBook},\cite{6DP}
\begin{equation}
\lbrack N]_{6}=\bigoplus_{\omega =N,N-2,...,0(1)}(\omega
,0,0)_{6}=\bigoplus_{i=0}^{<\frac{N}{2}>}(N-2i)_{6},  \label{U6O6}
\end{equation}
where $<\frac{N}{2}>=\frac{N}{2}$ if $N$ is even and $\frac{N-1}{2}$
if $N$ is odd. Furthermore, the following relation between the
quadratic \ Casimir operators $\overline{G_{3}}$ of
$\overline{SU(3)}$, $M^{2}$ of $\ SO(2)$\ and $C_{2}[O(6)]$  of
$O(6) $ holds \cite{6DP}:
\begin{equation}
C_{2}[O(6)]=2\overline{G_{3}}-\frac{1}{3}M^{2},  \label{Casimirs}
\end{equation}%
which means that the reduction from $O(6)$ to the rotational group
$SO(3)$ is carried out through the complementary groups $SO(2)$\ and
$\overline{SU(3)}$. As a consequence, the IR labels
$[f_{1},f_{2},0]_{3}$ of $SU(3)$ are determined by $(\omega )_{6}$
of $O(6)$\ and by the integer label $(\nu )_{2}$ of the associated
IR of $SO(2)$ i.e.
\begin{equation}
(\omega )_{6}=\bigoplus [f_{1},f_{2},0]_{3}\otimes (\nu )_{2}.
\label{o6su3}
\end{equation}%
Using the relation (\ref{Casimirs}) of the Casimir operators, for
their respective eigenvalues one obtains:
\begin{equation}
\omega (\omega +4)=\frac{4}{3}(f_{1}^{2}+f_{2}^{2}-f_{1}f_{2}+3f_{1})-\frac{%
\nu ^{2}}{3}.  \label{wfv}
\end{equation}
Thus (\ref{o6su3}) can be rewritten as \cite{6DP}
\begin{eqnarray*}
(\omega )_{6} &=&\bigoplus_{i=0}^{\omega }[\omega ,i,0]_{3}\otimes
(\omega
-2i)_{2}= \\
&&\bigoplus_{\nu =\omega ,\omega -2,...,0(1)}[\omega ,\frac{\omega -\nu }{2}%
,0]_{3}\otimes (\nu )_{2},
\end{eqnarray*}%
or in terms of the Elliott's notation $(\lambda ,\mu )$
\begin{equation}
(\omega )_{6}=\bigoplus_{\nu =\omega ,\omega
-2,...,0(1)}(\frac{\omega +\nu }{2},\frac{\omega -\nu }{2})\otimes
(\nu )_{2}.  \label{O6U3}
\end{equation}

Finally, the convenience of this reduction can be further enhanced
through the use of the standard rules (\ref{su3so3}) for the
reduction of the $\overline{SU(3)}\supset SO(3)$ chain.

\section{The boson condensate}

Usually, the condensate coherent state is defined in terms of the
'condensate boson' creation operator, which is a general linear
combination \cite{IBMQPT},\cite{DER}
\begin{equation}
B\equiv \alpha_{1}b_{1}+\alpha_{2}b_{2}+ \ldots
+\alpha_{M}b_{M},\label{CB}
\end{equation}
where in general the coefficients $\alpha_{i}$ are complex. Then the
(unnormalized) \emph{condensate} coherent state is
\cite{IBMQPT},\cite{DER}
\begin{align}
\mid N;\alpha_{1},\ldots,\alpha_{M} \ \rangle &\propto
(B^{\dagger})^N \mid 0 \ \rangle \notag \\
&=\left[\sum_{k} \alpha_{k}b^{\dagger}_{k}\right]^N \mid 0 \
\rangle, \label{BC}
\end{align}
where $\mid 0 \ \rangle$ is the boson vacuum. The condensate
coherent states (\ref{BC}) are often called "projective" CS. The
expectation value of a one-body or two-body operator with respect to
the condensate (\ref{BC}) can be deduced using arguments based upon
formal differentiation \cite{DER}.

Using the fact that the components of the two vector bosons ($p_{k}$
and $n_{k}$, respectively) form a six-dimensional vector, the
(unnormalized) CS for IVBM become
\begin{eqnarray}
\mid N;\xi,\zeta \ \rangle \propto \left[\sum_{k}
(\xi_{k}p_{k}^{\dagger}+\zeta_{k}n_{k}^{\dagger})\right]^N \mid 0 \
\rangle, \label{IVBMCS}
\end{eqnarray}
where $\xi_{k}$ and $\zeta_{k}$ are the components of
three-dimensional complex vectors. For static problems these
variables can be chosen real. The expression (\ref{IVBMCS}) defines
the so called projective realization of the CS for the fully
symmetric representation $[N]_{6}$ of $SU(6)$. We want to point out
that in contrast to the definition of the CS for IBM-2
\cite{PSDiep},\cite{PSIBM2a},\cite{PSIBM2b}, where the numbers of
protons, $N_{\pi}$, and neutrons, $N_{\nu}$, are separately
conserved, here only the total number of the two vector bosons
$N=N_{p}+N_{n}$ is a good quantum number. The parameters $\xi_{k}$
and $\zeta_{k}$ will determine the deformation of the nucleus in the
equilibrium state.

\section{Geometry}

The geometry can be chosen such that  $\overrightarrow{\xi}$ and
$\overrightarrow{\zeta}$  to span the $xz$ plane with the x-axis
along $\overrightarrow{\xi}$ and $\overrightarrow{\zeta}$ is rotated
by an angle $\theta$ about the out-of-plane y-axis,
$\overrightarrow{\xi} \cdot
\overrightarrow{\zeta}=r_{1}r_{2}\cos\theta$. In this way, the
condensate can be parametrized in terms of two real coordinates
$r_{1}$ and $r_{2}$ (the lengths of the two vectors), and their
relative angle $\theta$ ($r_{1},r_{2} \geq 0$  and $0 \leq \theta
\leq \pi$) in the following way
\begin{equation}
\mid N;r_{1},r_{2},\theta \ \rangle =
\frac{1}{\sqrt{N!}}(B^{\dagger})^{N}\mid 0 \ \rangle \label{BC2}
\end{equation}
with
\begin{equation}
B^{\dag}=\frac{1}{\sqrt{r_{1}^{2}+r_{2}^{2}}}\left[r_{1}
p^{\dagger}_{x} + r_{2}(n^{\dagger}_{x} \cos\theta +
n^{\dagger}_{z}\sin\theta)\right]. \label{BdagDec}
\end{equation}
The geometric properties of the ground states of nuclei within the
framework of the IVBM can then be studied by considering the energy
functional
\begin{equation}
E(N;r_{1},r_{2},\theta )=\frac{\langle
N;r_{1},r_{2},\theta|H|N;r_{1},r_{2},\theta \rangle}{\langle
N;r_{1},r_{2},\theta|N;r_{1},r_{2},\theta \rangle}. \label{Ener}
\end{equation}
By minimizing $E(N;r_{1},r_{2},\theta )$ (\ref{Ener}) with respect
to $r_{1}$, $r_{2}$, and $\theta$ , $\partial E/\partial r_{1}
=\partial E/\partial r_{2} = \partial E/\partial \theta = 0$, one
can find the equilibrium "shape" corresponding to any boson
Hamiltonian, $H$. This shape is in many cases rigid. However, there
are many situations in which the system is rather floppy and
undergoes a phase transition between two different shapes.

\section{Shape structure of the dynamical symmetries}

\subsection{The $U_{p}(3)\otimes U_{n}(3)$ limit}

\begin{figure}[h]\centering
\includegraphics[width=80mm]{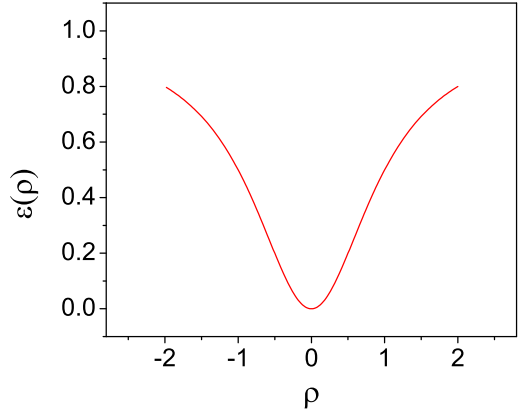}
\caption{(Color online) The scaled energy surface
$\varepsilon(\rho)$ in the $U_{p}(3)\otimes U_{n}(3)$ limit.}
\label{ESu3u3}
\end{figure}

We consider the Hamiltonian that is linear combination of first
order Casimirs of $U_{\tau}(3)$:
\begin{equation}
H_{I}=\varepsilon_{p} N_{p}+\varepsilon_{n} N_{n}. \label{Hv}
\end{equation}
The Hamiltonian (\ref{Hv}) can be rewritten in the form
\begin{equation}
H_{I}=\varepsilon_{p}N +\varepsilon N_{n}, \label{Hv2}
\end{equation}
where $\varepsilon=\varepsilon_{n}-\varepsilon_{p}$. The first term
in (\ref{Hv2}) can be dropped since it does not contribute to the
energy surface. Thus, the Hamiltonian determining the properties of
the system in the $U_{p}(3)\otimes U_{n}(3)$ limit is just
\begin{equation}
H_{I}=\varepsilon N_{n}. \label{Hvib}
\end{equation}
The expectation value of (\ref{Hvib}) with respect to (\ref{BC2})
gives the energy surface
\begin{equation}
\frac{\langle N;r_{1},r_{2}|H_{I}|N;r_{1},r_{2}\rangle}{\langle
N;r_{1},r_{2}|N;r_{1},r_{2}\rangle}=\varepsilon
\frac{Nr^{2}_{2}}{r^{2}_{1}+r^{2}_{2}} . \label{Ev}
\end{equation}
We see that the energy surface depends on the two parameters $r_{1}$
and $r_{2}$ and is $\theta-$independent. In order to simplify the
analysis we introduce a new dynamical variable $\rho=r_{2}/r_{1}$ as
a measure of "deformation", which together with the parameter
$\theta$ determine the corresponding "shape". Thus, the expression
(\ref{Ev}) becomes
\begin{equation}
E(N;\rho)=\varepsilon \frac{N\rho^{2}}{1+\rho^{2}} , \label{Evib}
\end{equation}
which has a minimum at $\rho_{0}=0$. It corresponds to a spherical
shape (vibrational limit). The scaled energy
$\varepsilon(\rho)=E(N;\rho)/\varepsilon N$ in the $U_{p}(3)\otimes
U_{n}(3)$ limit is given in Figure \ref{ESu3u3}. The inclusion of
higher-order terms in $N_{p}$ and $N_{n}$ will give rise to an
anharmonicity.

\subsection{The $O(6)$ limit}

The Hamiltonian describing the $O(6)$ (or $\gamma-$unstable)
properties can be written down through the $O(6)$ pairing operator
$P^{\dag}=\frac{1}{2}(p^{\dag} \cdot p^{\dag}-n^{\dag} \cdot
n^{\dag})$ in the following form
\begin{equation}
H_{II}=\frac{4k'}{N-1}P^{\dag}P. \label{HO6}
\end{equation}
In (\ref{HO6}) the $P^{\dag}P$ operator is used instead of the
quadratic Casimir operator $C_{2}[O(6)]$ of $O(6)$ because of their
linear dependence, i.e. $C_{2}[O(6)]=-P^{\dag}P + N(N+4)$.

\begin{figure}[h]\centering
\includegraphics[width=80mm]{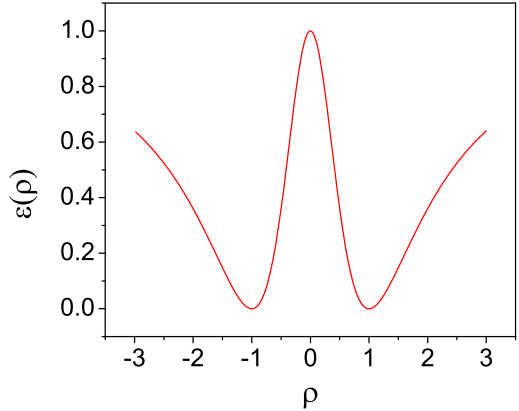}
\caption{(Color online) The scaled energy surface
$\varepsilon(\rho)$ in the $O(6)$ limit.} \label{ESo6}
\end{figure}

Taking the expectation value of (\ref{HO6}) one obtains the energy
surface
\begin{equation}
E(N;\rho)=k'N\left[\frac{1-\rho^{2}}{1+\rho^{2}}\right]^{2},
\label{EO6}
\end{equation}
which does not depend on $\theta$ ($\theta-$unstable) and has a
minimum at $\rho_{0} \neq 0$ ($|\rho_{0}|= 1$). It corresponds to a
deformed "$\gamma-$unstable" (in IBM terms) rotor. As it was
mentioned, there are two $O_{\pm}(6)$ algebras that are isomorphic
and have the same eigenspectrum but differ through phases in the
wave functions resulting into different energy surfaces. The energy
surface (\ref{EO6}) corresponds to the $O_{-}(6)$ limit. The other,
$O_{+}(6)$, limit is not physically important since its energy
surface is just a constant. The scaled energy surface
$\varepsilon(\rho)$ in the $O_{-}(6)$ limit is given in Figure
\ref{ESo6}.

\subsection{The $\overline{SU_{\pm}(3)}$ limit}

\begin{figure}[h]\centering
\includegraphics[width=80mm]{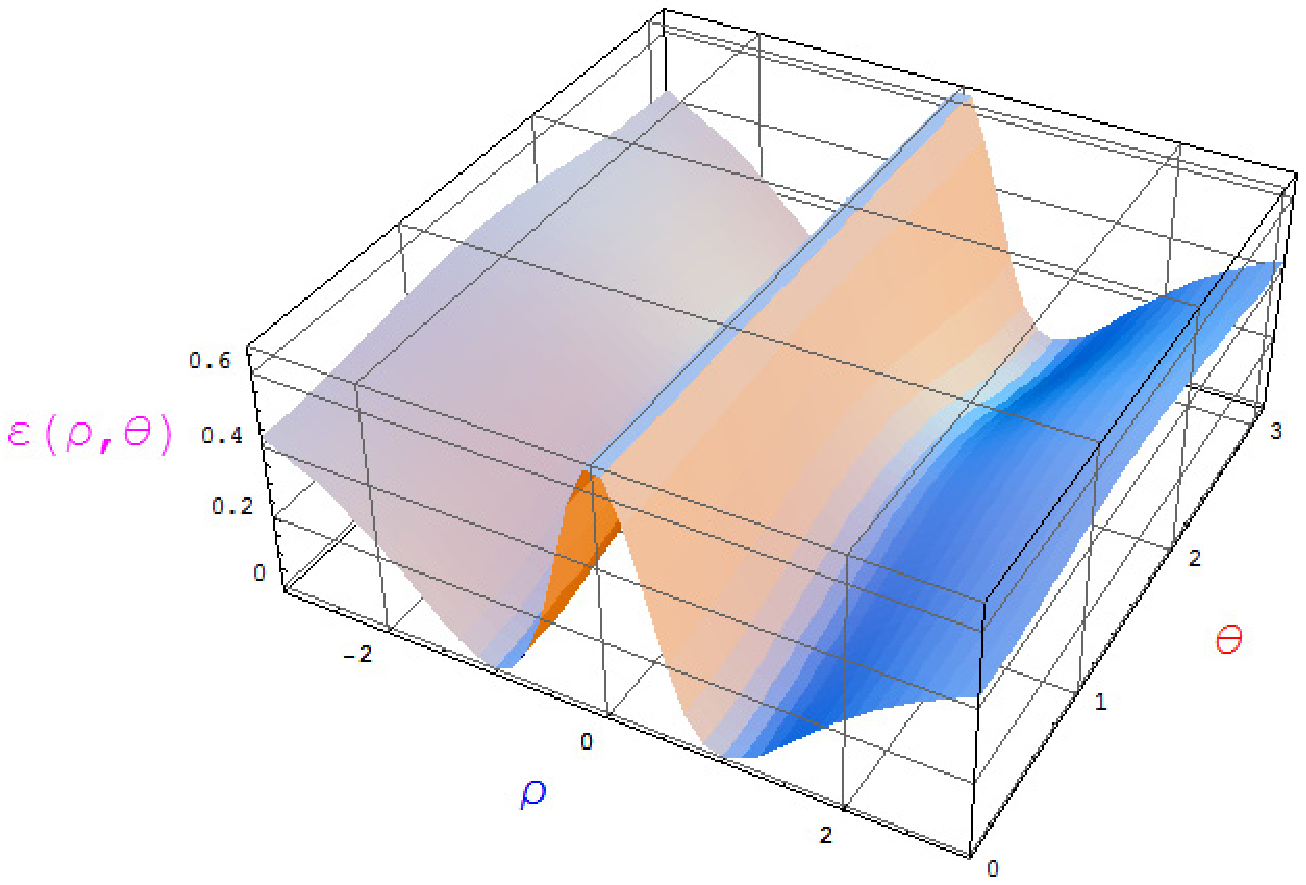}
\caption{(Color online) The scaled energy surface
$\varepsilon(\rho,\theta)$ in the $\overline{SU_{+}(3)}$ limit.}
\label{ESsu3bp}
\end{figure}

\begin{figure}[h]\centering
\includegraphics[width=80mm]{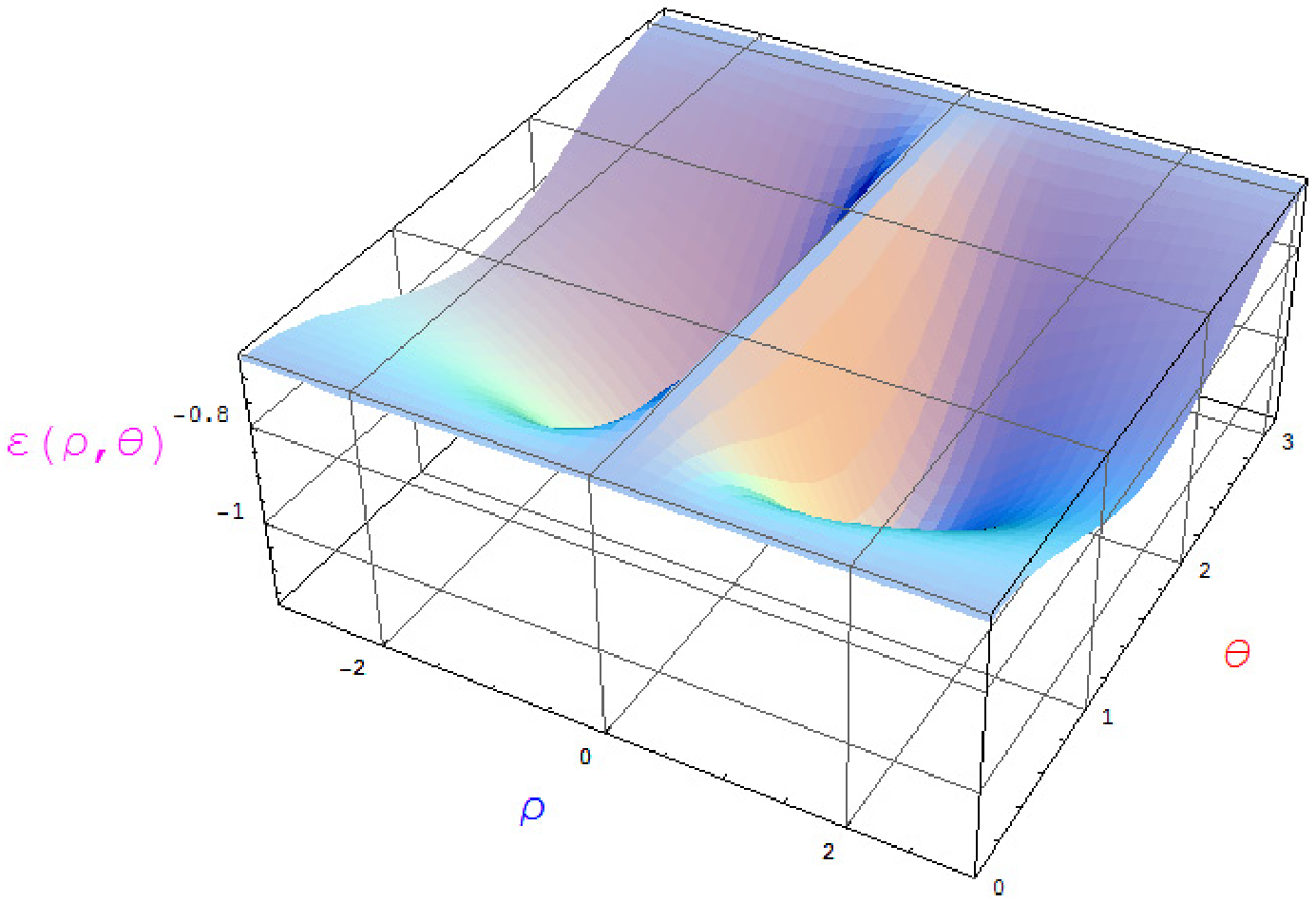}
\caption{(Color online) The scaled energy surface
$\varepsilon(\rho,\theta)$ in the $\overline{SU_{-}(3)}$ limit.}
\label{ESsu3bm}
\end{figure}

In this case we study the Hamiltonian
\begin{equation}
H_{III}=-\frac{k}{(N-1)}\overline{G_{3}}, \label{HSU3B}
\end{equation}
where $\overline{G_{3}}$ is given by (\ref{C2SU3B}). Note that the
operators $B^{(\omega)}_{ij}$ ($\omega=\pm 1$) entering in
$\overline{G_{3}}$ generate the two distinct algebras
$\overline{SU_{+}(3)}$ and $\overline{SU_{-}(3)}$, respectively. The
energy surface corresponding to the Hamiltonian (\ref{HSU3B}) is
\begin{equation}
E(N;\rho,\theta)=-kN\left[\frac{2\rho^{2}(\cos^{2}\theta+3)}{3(1+\rho^{2})^{2}}
+\frac{2}{3}\right], \label{ESU3Bp}
\end{equation}
for $\overline{SU_{+}(3)}$, and
\begin{equation}
E(N;\rho,\theta)=-kN\left[\frac{2\rho^{2}\sin^{2}\theta}{(1+\rho^{2})^{2}}+\frac{2}{3}\right],
\label{ESU3Bp}
\end{equation}
for $\overline{SU_{-}(3)}$ algebra, respectively. We plot the scaled
energy surfaces corresponding to the two limits under consideration
in Figures \ref{ESsu3bp} and \ref{ESsu3bm}, respectively. From the
figures one can see that for both cases the global minimum occurs at
$\rho_{0} \neq 0$ ($|\rho_{0}|=1$) and $\theta_{0} =0^{0}$ or
$\theta_{0} = 90^{0}$. The fact that the deformation parameter
$|\rho|=1$ means that we have equal deformations of the $p-$ and
$n-$boson subsystems which ratio is given by
$\rho^{2}=\frac{r^{2}_{2}}{r^{2}_{1}}=\frac{N_{n}}{N_{p}}$.
Similarly, the equilibrium configurations for the combined
proton-neutron $SU(3)$ and $SU^{\ast}(3)$ phase shapes in IBM-2 are
obtained for $\beta_{\pi}=\beta_{\nu}$
\cite{PSDiep},\cite{PSIBM2a},\cite{PSIBM2b}.

\subsection{The $SU(3)\otimes U_{T}(2)$ limit}

This limit can be studied through the Hamiltonian
\begin{equation}
H_{IV}=-\frac{k}{(N-1)}K_{3}, \label{HSU3}
\end{equation}
where the second order Casimir operator $K_{3}$ of $SU(3)$ is given
by (\ref{C2SU3}). The expectation value of (\ref{HSU3}) with respect
to (\ref{Ener}) gives the following energy surface
\begin{equation}
E(N;\rho,\theta)= -kN\left[\frac{(1+2\rho^{2}\cos^{2}\theta
+\rho^{4})}{(1+\rho^{2})^{2}}-\frac{2}{3}\right]. \label{ESU3}
\end{equation}
For positive values of the parameter $k > 0$ one obtains an
oscillator in the relative angle $\theta$, which has the equilibrium
at $\theta_{0}=0$. To show this one needs to consider a more general
semi-classical analysis of the classical limit of H in which the
complex coherent state parameters are used.

\begin{figure}[h]\centering
\includegraphics[width=80mm]{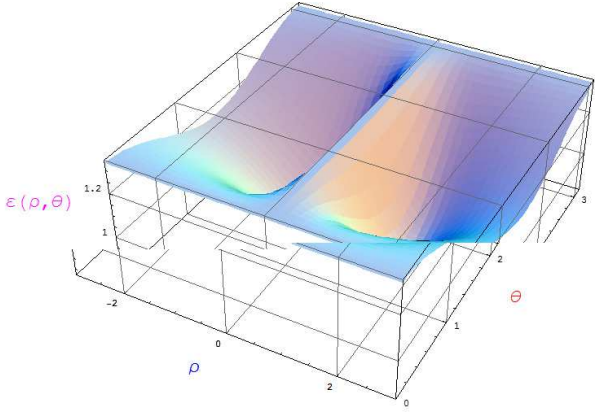}
\caption{(Color online) The scaled energy surface
$\varepsilon(\rho,\theta)$ in the $SU(3)\otimes U(2)$ limit.}
\label{ESsu3}
\end{figure}

For negative values of the parameter $k < 0$, the energy surface in
the $SU(3)\otimes U_{T}(2)$ limit is given in Figure \ref{ESsu3}.
The deformed phase shape is determined through the equilibrium
values $|\rho_{0}|=1$ and $\theta_{0} = 90^{0}$, respectively.

\section{Relation with the Bohr triaxial variable}

In order to clarify to what geometry correspond the energy surfaces
(potentials) obtained in the preceding sections, we consider the
relation of $\theta$ with the standard triaxial "shape" parameter
$\gamma$ in the collective model. A relation between standard Bohr
collective model shape variables used to describe the deformation of
the collective motion and the shape parameters in the intrinsic
state of IVBM can be obtained by calculating the expectation value
of the quadrupole moments of the corresponding dynamical symmetry
limit with respect to the IVBM coherent state. In the intrinsic
state of IVBM, the effective $\gamma_{eff}$ deformation can be
defined in the usual way as \cite{BM}:
\begin{equation}
\tan \gamma_{eff} =\sqrt{2}\frac{\langle Q_{2} \rangle}{\langle
Q_{0} \rangle}, \label{Gamma}
\end{equation}
where by $\langle Q_{\mu} \rangle$ is denoted the expectation value
of the $\mu$th component of the quadrupole operator defined for the
dynamical symmetry limit under consideration.

For the $SU(3)\otimes U_{T}(2)$ limit one obtains:
\begin{equation} \tan \gamma_{eff} =
\frac{\sqrt{3}(\rho^{2}\cos^{2}\theta+1)}{\left[\frac{\rho^{2}}{2}(-3\cos2\theta+1)-1\right]}
. \label{GammaTita1}
\end{equation}
The expression (\ref{GammaTita1}) gives a relation between the
"projective" IVBM CS deformation parameters $\{\rho,\theta\}$ and
the standard collective model parameter $\gamma_{eff}$, determining
the triaxiality of the nuclear system. As we saw in the previous
sections, for the equilibrium values of the IVBM shape parameters
$|\rho_{0}|=1$ and $\theta_{0}=90^{0}$ in the $SU(3)\otimes
U_{T}(2)$ limit one obtains $\gamma_{eff}=60^{0}$. This corresponds
to an oblate axial configuration of the compound $pn-$system.

Analogously, one can show that the $\overline{SU_{+}(3)}$ and
$\overline{SU_{-}(3)}$ phase shapes correspond to an oblate and
prolate axial deformation, respectively, of the two-fluid nuclear
system.

Finally, we note that the constraint $|\rho_{0}|=1$ is valid only
for the limiting cases of dynamical symmetry limits. Thus for
intermediate situations, one can get values for $\gamma_{eff}$
different from $0^{0}$ or $60^{0}$, corresponding to prolate or
oblate deformed axial configuration of the nuclear system.

\section{The generalized IVBM Hamiltonian and its phase diagram}

\begin{figure}[h]\centering
\includegraphics[width=80mm]{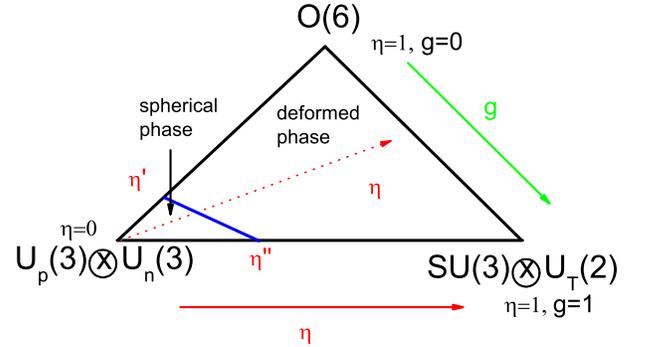}
\caption{(Color online) Phase diagram of IVBM. The corners of the
triangle correspond to dynamical symmetries.} \label{pyramid}
\end{figure}

All physically interesting Hamiltonians can be combined into a
single Hamiltonian that keeps all main ingredients of the considered
limits. It is convenient again to scale the parameters of the
two-body terms by $(N-1)$ and to consider the following Hamiltonian:
\begin{equation}
H=(1-\eta) N_{n} +\frac{\eta}{N-1}\left[gK_{3} + (1-g)P^{\dag}P
\right], \label{Hcom}
\end{equation}
The three terms in (\ref{Hcom}) correspond to the three dynamical
symmetries: $U_{p}(3)\otimes U_{n}(3)$, $O(6)$ and $SU(3)\otimes
U_{T}(2)$. We have two control parameters: $\eta$ and $g$ and hence
the resulting phase diagram is two dimensional one. The phase shape
diagram corresponding to the IVBM Hamiltonian (\ref{Hcom}) can be
depicted as a triangle as shown in Figure \ref{pyramid} with each
corner denoting a dynamical symmetry. For $\eta=0$ one obtains the
$U_{p}(3)\otimes U_{n}(3)$ or the \emph{vibrational} limit; for
$\eta=1$ one encounters the two limiting cases of \emph{deformed}
shapes discussed above: $g=0$ ($O(6)-\gamma$-unstable rotor) and
$g=1$ ($SU(3)-$axial rotor).

In some cases, the quantum phase transitions can take place between
different ground state configurations or "shapes" of the system,
occurring at zero temperature as a function of the corresponding
control parameter. The order of the phase transitions may be
determined with the standard approach (see, for instance, Ref.
\cite{IBM}). Here the phase transitions will be studied by analyzing
the behavior of the order parameter as a function of the control
parameter (Landau's approach).

\begin{figure}[h]\centering
\includegraphics[width=80mm]{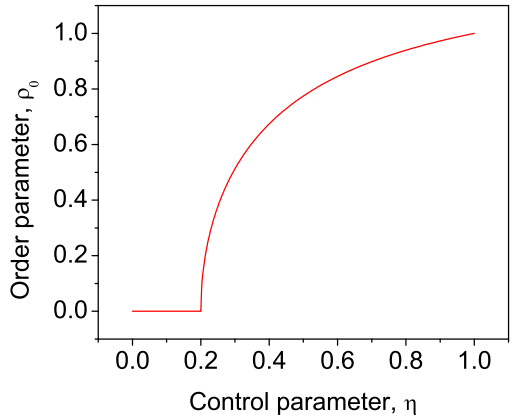}
\caption{(Color online) Classical order parameter $\rho_{0}$ as a
function of the control parameter $\eta$ for $g=0$.} \label{OPa}
\end{figure}

It can be shown that along the leg $U_{p}(3)\otimes U_{n}(3)-O(6)$ a
second order phase transition is observed at the critical value of
the parameter $\eta'_{c}=0.2$. For $\eta < \eta'_{c}$ an equilibrium
spherical shape ($\rho_{0}=0$) is obtained, while for $\eta >
\eta'_{c}$ the equilibrium shape is deformed ($|\rho_{0}|=1$). The
behavior of $\rho_{0}$ as a function of $\eta$ is shown in Figure
\ref{OPa}, displaying the typical behavior of a second-order
transition \cite{PT}. This confirms the generic statement that for
models based on $U(n_{1}+n_{2})$, the phase transition between the
two phases $U(n_{1})\otimes U(n_{2})$ and $O(n_{1}+n_{2})$ is of
second order \cite{TLPM}.

\begin{figure}[h]\centering
\includegraphics[width=80mm]{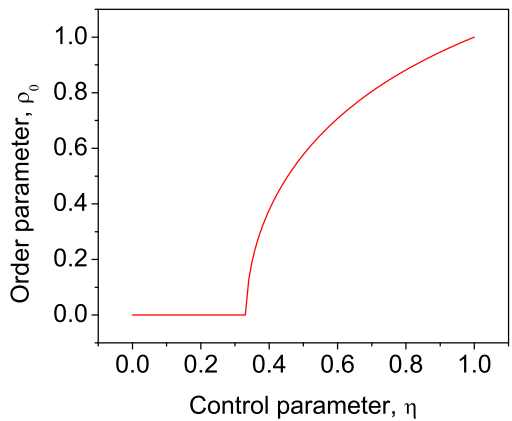}
\caption{(Color online) Classical order parameter $\rho_{0}$ as a
function of the control parameter $\eta$ for $g=1$.} \label{OPb}
\end{figure}

A second order phase transition between spherical and deformed axial
phase shapes (the $U_{p}(3)\otimes U_{n}(3)-SU(3)\otimes U_{T}(2)$
leg) is observed at $\eta''_{c}=0.33$. In Figure \ref{OPb} the
behavior of the order parameter $\rho_{0}$ as a function of $\eta$
is shown for the $U_{p}(3)\otimes U_{n}(3)-SU(3)\otimes U_{T}(2)$
transition. From Figure \ref{OPb} one can see again the typical
behavior of a second-order transition \cite{PT}.

Taking the expectation value of the more general Hamiltonian
(\ref{Hcom}) one obtains its classical limit (Landau potential) in
the following form:
\begin{equation}
\varepsilon(\rho,\theta)=
\frac{\varepsilon_{0}+A(\theta)\rho^{2}+C\rho^{4}}{(1+\rho^{2})^{2}},
\label{LP}
\end{equation}
where  $A(\theta)=1+\eta\left[2g(1+\cos^{2}\theta)-3\right]$, $C=1$,
and $\varepsilon_{0}=\eta$. The potential (\ref{LP}) has a generic
Landau-like form, except that the cubic term is missing. Note also
that, in contrast to the traditional Landau theory, the denominator
in (\ref{LP}) ensures a finite potential when $\rho \rightarrow
\infty$. The minimization in the variable $\theta$ can be performed
separately since the dependence of the potential (\ref{LP}) is only
through the $\cos^2\theta$. This potential has a minimum either at
$\theta_{0}=0^{0}$ or $\theta_{0}=90^{0}$ and setting to these
values we can study only the $\rho$ dependence. The global minimum
of the potential (\ref{LP}) can be found by its minimization with
respect to $\rho$. If such a minimum occurs at ($\rho = 0$) one has
the higher (spherical) symmetry and for ($\rho \neq 0$), the lower
(deformed) symmetry. The minimization gives two solutions
\begin{equation}
\rho_{0}=0,  \label{sphersol}
\end{equation}
and (setting $\theta_{0}=90^{0}$)
\begin{equation}
\rho_{0}=\pm \sqrt{\frac{1-5\eta+2g\eta}{-1-3\eta+2g\eta}}.
\label{defsol}
\end{equation}
When $g=0$ or $g=1$, we recover the familiar results for the
critical point values $\eta_{c}=0.2$ or $\eta_{c}=0.33$ at which a
phase transition is observed from spherical to $\gamma-$unstable
deformed or from spherical to axially deformed phase shape. The
critical value $\eta_{c}$ at which the phase transition appears
moves from $0.2$ to $0.33$ with the increase of $g$ from $0$ to $1$.

\begin{figure}[h]\centering
\includegraphics[width=80mm]{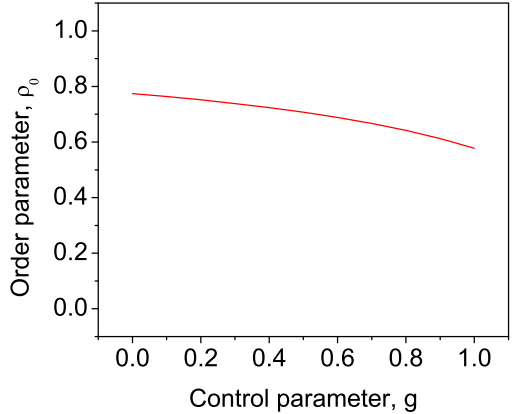}
\caption{(Color online) Classical order parameter $\rho_{0}$ as a
function of the control parameter $g$ for $\eta=0.5$.} \label{OP3}
\end{figure}

For $\eta=1$, we obtain a constant value of $\rho_{0}=1$ as a
function of $g$ (the $O(6)-SU(3)\otimes U_{T}(2)$ leg of Fig.
\ref{pyramid}), which indicates that no phase transition occurs. For
a fixed value of $\eta < 1$ the equilibrium value of $\rho_{0}$
decreases smoothly. In Figure \ref{OP3}, the behavior of the order
parameter $\rho_{0}$ as a function of the control parameter $g$ is
shown for $\eta=0.5$.

\begin{figure}[h]\centering
\includegraphics[width=80mm]{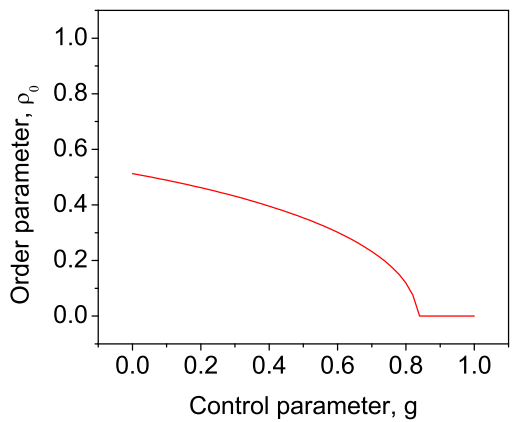}
\caption{(Color online) Classical order parameter $\rho_{0}$ as a
function of the control parameter $g$ for $\eta=0.3$.} \label{OP4}
\end{figure}

For $\eta=0.33$ a second order phase transition occurs from deformed
to spherical phase shape at the critical point $g=1$, which moves to
the lower values of $g$ with the further decrease of $\eta$ ($>
0.2$). This phase transition as a function of $g$ disappears at
$\eta = 0.2$. In Figure \ref{OP4}, the behavior of the order
parameter $\rho_{0}$ as a function of the control parameter $g$ is
shown for $\eta=0.3$.

We note that due to the missing of the cubic term in (\ref{LP}) the
first order phase transitions do not appear. Hence the
spherical-deformed second order phase transition occurs along the
continuous line $\eta'-\eta''$, but not at isolated points. Indeed,
from Eq.(\ref{defsol}) one easily obtains the critical line equation
\begin{equation}
\eta=\frac{1}{5-2g},  \label{critline}
\end{equation}
which reduces to the two limiting critical points $\eta'=0.2$ (g=0)
and $\eta''=0.33$ (g=1), respectively.

\section{Summary}

In the present paper, the geometrical analysis of the different
dynamical symmetries of the two-fluid IVBM with the $U(6)$ as a
dynamical group is carried out by means of coherent state method.
The latter allows the calculation of the classical limit of the
Hamiltonians corresponding to different dynamical symmetries in
terms of appropriately chosen classical (geometrical) variables
representing the boson degrees of freedom. The different dynamical
symmetries correspond to qualitatively distinct ground state
equilibrium configurations, which constitute the phases of the
system. Thus, the dynamical symmetries are the ones that determine
the structure of the corresponding phase diagram.

We have studied the phase structure of IVBM and three phase shapes
corresponding to its three dynamical limits have been obtained:
spherical, $U_{p}(3)\otimes U_{n}(3)$, $\gamma-$unstable, $O(6)$,
and axially deformed shape, $SU(3)\otimes U_{T}(2)$. Further, the
ground state quantum phase transitions between different phase
shapes, corresponding to the different dynamical symmetries and
mixed symmetry case, are investigated theoretically using one or
(for the more general IVBM Hamiltonian (\ref{Hcom})) two control
parameters in the Hamiltonian. The latter drive the system in
different phases. It is shown that a second order phase transition
occurs from spherical to deformed phase shape.

In conclusion, we concern the question of the possible physical
interpretation of the introduced classical variables. In general,
there is a variety of ways in which the geometrical variables can be
defined. Under some constraints different realizations of the CS can
be related to each other. In practice, it turns out that the most
convenient are the "projective" coherent states, which for the IVBM
were defined by (\ref{IVBMCS}). The two set of dipole variables
$\{\xi_{k}\}$ and $\{\zeta_{k}\}$ entering in (\ref{IVBMCS}) can be
related with the quadrupole-octupole deformations realized in many
regions of nuclear chart. The presence of octupole deformation
causes a shift of the nuclear center of mass which must be balanced
by addition of a dipole deformation, which in lowest order is
proportional to the product of quadrupole and octupole deformations.

In our opinion the physical meaning should be associated with the
physical quantities which can be constructed from the "projective
variables" rather than these variables themselves. This can be
easily seen if the "algebraic" CS \cite{IBM} are used. The
introduction of the algebraic CS (and the respective "algebraic
geometrical variables") is based on the precise mathematical
procedure related with the theory of the coset spaces, which allows
one to attach a geometrical space to a certain algebra g.
Geometrical variables are then associated with the elements of the
considered coset space. In this regard, the IVBM turns out to be a
particular case of a more general class of algebraic models, called
two-level pairing models \cite{TLPM}. An associated geometry for the
two-level pairing models is defined through the coset space
$U(n_{1}+n_{2})/U(n_{1}) \otimes U(n_{2})$ (where $n_{1}=2L_{1}+1$
and $n_{2}=2L_{2}+1$), which obviously generalizes the coset space
of the usual two-level boson models for which $n_{1}=1$ (i.e. one of
the bosons is a scalar boson, $s$). In our case we have
$n_{1}=n_{2}=3$ which leads to a much richer phase space in
$9-$dimensions, including in addition to the quadrupole also
monopole and dipole collective degrees of freedom.

Finally, one might relate the "projective variables" with the
cluster degrees of freedom in nuclei, connected with the relative
motion of the clusters. The cluster degrees of freedom as well as
the octupole ones play an important role in the description of the
negative parity states in nuclei.

\section*{Acknowledgments}

This work was supported by the Bulgarian National Foundation for
scientific research under Grant Number DID-$02/16/17.12.2009$. The
author is grateful to Alberto Ventura and Ana Georgieva for reading
the manuscript.

\end{document}